\documentstyle[preprint,revtex]{aps}
\tightenlines
\begin{document}
\draft
\preprint{DUKE-TH-92-38}
\vskip3mm
\preprint{Revised 11/7/92}
%
%
\begin{title}
{\bf Vacuum Polarization}\\
{\bf and the Electric Charge of the Positron}
\end{title}
\author{Berndt M\"uller$^*$ and Markus H. Thoma$^{\dagger}$}
\begin{instit}
Department of Physics, Duke University, Durham, NC 27708-0305
\end{instit}
\vspace{0.1in}
%
\begin{abstract}
We show that higher-order vacuum polarization would contribute
a measureable net charge to atoms, if the charges of electrons
and positrons do not balance precisely. We obtain the limit
$|Q_e+Q_{\bar e}| < 10^{-18} e$ for the sum of the charges
of electron and positron. This also constitutes a new bound on
certain violations of PCT invariance.
\end{abstract}
%
\baselineskip 1cm

In a recent Letter \cite{HD92} Hughes and Deutch discussed the possibility
that the charges of positrons and antiprotons may not be exactly opposite
to those of electrons and protons. Whereas the equality in magnitude of the
charges of electrons and protons is known to the extreme accuracy
\cite{DK73,MM84}
\begin{equation}
|Q_e+Q_p| < 10^{-21} e ,
\end{equation}
the equality in magnitude of the charges of electrons and positrons is
much more difficult to study directly. After
reviewing the available body of evidence, Hughes and Deutch conclude
that the present limit on the net neutrality of an electron-positron
pair is
\begin{equation}
|Q_e+Q_{\bar e}| < 4\times 10^{-8} e .
\end{equation}

Here we would like to point out that there exist far more stringent
bounds on this quantity from indirect sources. Our argument is based
on the fact that the vacuum polarization in
heavy atoms contains an equal number of electrons and positrons
and hence would contribute to the overall charge of an atom, if the
charges of electrons and positrons do not balance each other
precisely. This reasoning is closely related to the observation first
made by Morrison \cite{Mo58} and Schiff \cite{Sc59}, that the
equality of the gravitational masses of electrons and positrons
is probed to about 1 percent accuracy by the fact that the
contribution of vacuum polarization to the mass of an atom does
not lead to a violation of the equivalence principle.

As we will show below, this argument is much
more powerful in the case of the electric charge.
In fact, our bound would be even more precise, would it not
be for the necessity of charge renormalization. Since the amount
of charge contained in the lowest order (in $Z\alpha$, where $Z$ is
the nuclear charge) vacuum polarization is directly proportional to
the source charge of the Coulomb field, the net vacuum polarization
charge to this order can be absorbed in the {\it renormalized} charge
of the source, rendering it effectively unobservable. This reasoning does
not apply to higher orders in $Z\alpha$ of the atomic vacuum polarization,
which do not contribute to charge renormalization.

If the charges of electrons and positrons are not opposite and equal,
the first nonvanishing contribution to the overall charge of an atom
by the vacuum polarization would come in order $(Z\alpha)^2$.
According to Furry's theorem \cite{Fu37}, this
order normally vanishes identically due to the invariance of quantum
electrodynamics (QED) against charge conjugation (C-invariance).
However, if $Q_e$ and $Q_{\bar e}$ do not balance
each other, this would imply a violation of C-invariance and hence
invalidate Furry's theorem.

It is not clear that a completely consistent quantum field theory
of QED without C-invariance can be constructed, but for our purposes
it is sufficient to consider an effective theory that is consistent
at the one-loop level. This is provided by the Lagrangian
\begin{equation}
L = \bar\psi(i\gamma^\mu\partial_\mu - m)\psi
+[Q_e\bar\psi_e\gamma^\mu\psi_e
+Q_{\bar e}\bar\psi_{\bar e}\gamma^\mu\psi_{\bar e}
+\tilde{Q}(\bar\psi_e\gamma^\mu\psi_{\bar e} +
\bar\psi_{\bar e}\gamma^\mu\psi_e)]A_\mu ,
\end{equation}
where $\psi_{e/\bar e}=P_\pm\psi$ denotes the Dirac field projected
on positive and negative energies, respectively, $Q_e$ and $Q_{\bar e}$
are the charges of electron and positron, and $\tilde{Q}$
denotes the coupling constant associated with pair creation. From
the success of QED precision measurements we know that
$Q_e\equiv -e\approx -Q_{\bar e}$ at least to within $10^{-8}$ \cite{HD92}
and $\tilde{Q}\approx -e$ to within $10^{-3}$ \cite{W87}.

In addition to C-invariance, the Lagrangian (3) breaks gauge
and PCT invariance. The former expresses the fact that
charge conservation is violated if one assignes unequal charges
in magnitude to electron and positron, but allows for pair annihilation
into a neutral photon. The violation of PCT invariance is reconciled
with the Pauli-L\"uders theorem \cite{PL55} by noting that the
projection operators $P_{\pm}$ appearing in (3) are nonlocal.
They are given by:
\begin{equation}
(P_{\pm}\psi)(\vec{x},t) =  \mp\gamma^0{m\over 4\pi^2} \int d^3x'
\frac{K_2(m|\vec{x}-\vec{x'}|)}{|\vec{x}-\vec{x'}|^2} \psi(\vec{x'},t)
+{1\over 2m}(\pm i\vec{\gamma}\cdot\nabla +m)\psi(\vec{x},t),
\end{equation}
which is nonlocal on the scale of the electron Compton wavelength.
Although the breaking of gauge and PCT invariance may seem
unattractive, it is unavoidable if one wants to construct
a low-energy effective theory describing particles and antiparticles
with unequal opposite charges.

We now apply the Lagrangian (3) to the calculation of order
$\alpha(Z\alpha)^2$ vacuum polarization in atoms, which is the lowest
order where a nonvanishing $(Q_e+Q_{\bar e})$ would contribute.
The relevant Feynman diagrams describing the contribution of vacuum
polarization to Rutherford scattering on a nucleus are shown in Figure 1.
Intuitively, they correspond to scattering on the virtual
positrons (a) and electrons (b) in the polarization cloud around the
nucleus.  There is a time ordering $(x_0>y_0)$ assumed,
which is imposed by the nucleus.
Therefore arrows pointing up represent electron propagators $S_e(x-y)=\theta
(x_0-y_0)\> S^+(x-y)$, whereas arrows pointing down correspond to positron
propagators $S_{\bar e}(y-x)=-\theta (x_0-y_0)\> S^-(y-x)$. Here the
propagators $S^\pm$ are related to the Feynman propagator by \cite{MS84}
\begin{equation}
S_F(x-y) = \theta (x_0-y_0)\> S^+(x-y)-\theta (y_0-x_0)\> S^-(x-y) .
\end{equation}
After Fourier transformation we obtain
\[
S_e(p)=\frac {\gamma _0E_p-\vec\gamma\cdot\vec p+m}{2E_p}\>
\frac{1}{p_0-E_p+i\epsilon}, \nonumber
\]
\begin{equation}
S_{\bar e}(-p)=\frac{\gamma _0E_p+\vec\gamma\cdot\vec p-m}{2E_p}\>
\frac{1}{p_0+E_p-i\epsilon},
\end{equation}
where $E_p=\sqrt{\vec p\, ^2+m^2}$. These propagators together with the
vertices, modified by the coupling constants $Q_e$, $Q_{\bar e}$, and
$\tilde Q$ respectively, define the Feynman rules which have to be used in
Figure 1.
The contribution of the two diagrams to the scattering matrix
is strictly proportional to the charge imbalance:
\begin{eqnarray}
\Delta S_{\rm fi} &=& (Q_e+Q_{\bar e})Q{\tilde Q}^2
\int {d^4q\over (2\pi )^4} {d^4p\over (2\pi )^4}
A_{\mu}(q)A_{\nu}(q')D(q+q')\;
\bar{u}_{\rm f}\gamma_{\lambda}u_{\rm i}\; \nonumber\\
&&\hskip5cm
\hbox{tr}[S_{\bar e}(p+q')\gamma^{\nu}S_e(p)\gamma^{\mu}S_{\bar e}(p-q)
\gamma^{\lambda}] ,
\end{eqnarray}
where $A_{\mu}(q)$ represents the electromagnetic potential generated
by the nucleus and $Q$ denotes the charge of the scattering
particle. The loop integral over $p$ in (7) is superficially
linearly divergent, but is actually finite due to cancellation of
the leading orders in $p$. We also note that, in contrast to the
fourth-order contribution to vacuum polarization, there is no need for
a finite subtraction \cite{fn}.

For our purposes it is sufficient to
consider the limit of forward scattering ($q+q'=0$) in
the nonrelativistic limit, where only the time-like components
($\mu=\nu=\lambda=0$) contribute. Then it is easy to see that
the effect of (7) on the scattering amplitude corresponds to the
presence of an additional charge
\begin{equation}
\Delta Ze = (Q_e+Q_{\bar e})Q(Ze{\tilde Q})^2 \int {d^3q\over (2\pi)^3}
{F(\vec q\, ^2)^2\over \vec q\, ^4} \int {d^3p\over (2\pi)^3}
\frac{E_pE_{p-q} - E_p^2+\vec p\cdot \vec q}{E_pE_{p-q}(E_p+E_{p-q})^2}
\end{equation}
surrounding the nucleus. Here $F(q^2)$ is the nuclear elastic form
factor and $E_{p-q}=\sqrt{(\vec p-\vec q\, )^2+m^2}$. Since we are not
interested in extreme precision, we simply cut off the $q$-integration at
the inverse nuclear radius $R$ and evaluate the integrals in (8)
to leading order in the cut-off. We also set ${\tilde Q}=-e$.
The result is:
\begin{equation}
\Delta Ze \approx (Q_e+Q_{\bar e}){2Z^2\alpha^2\over 3\pi^2}
\left[ \ln\Bigl({1\over mR}\Bigr)+c\right] ,
\end{equation}
where the constant $c$ depends on the details of the nucleon form factor
and can be neglected for our purpose.
For a heavy atom, such as lead ($Z=82, R=7$ fm), we find
$\Delta Ze \approx {1\over 10}(Q_e+Q_{\bar e})$. With the limit (1)
on the apparent residual charge of the atom per proton, $\Delta Z/Z$,
we obtain the bound
\begin{equation}
|Q_e+Q_{\bar e}| < 10^{-18} e .
\end{equation}
Because the net vacuum polarization charge is quadratic
in the nuclear charge $Z$, it is impossible to simultaneously
adjust the electron-positron and electron-proton charge differences
such that all atoms are neutral, without satisfying the
bound (10). Since the momentum integrations in (8) involve only
momenta up to $R^{-1}$, and the structure of QED has been tested
to very high precision over that range, we believe that
our result is essentially model-independent.
Because our effective Lagrangian (3) breaks PCT-invariance, the
bound (10) can also be taken as a new test of PCT symmetry, which
is better by a factor 4 than the limit derived from the neutral
kaon system \cite{C90}, but tests a different mode of PCT symmetry
breaking.

In conclusion, we have shown that the existing limit on violations
of the neutrality of atoms sets a very stringent limit on the
opposite equality of electron and positron charge, if one
considers the second-order vacuum polarization, which normally
vanishes due to Furry's theorem. It is unlikely that direct
experimental tests can improve on this bound soon.

{\it Acknowledgments:} This work has been supported in part
by the U.S. Department of Energy (Grant DE-FG05-90ER40592).
We thank A. Sch\"afer for useful discussions on PCT invariance.

\figure{Diagrams contributing to second-order (in $Z\alpha$) vacuum
polarization correction to the forward Coulomb scattering cross
section on an atomic nucleus. Diagram (a) is proportional to the
positron charge $Q_{\bar e}$ and (b) to the electron charge $Q_e$.}



\begin{references}
\bibitem{fn}
	This can be seen in two ways: (a) No third-order term in the
	external field arises in Pauli-Villars regularization. (b)
	The only possible Lorentz invariant subtraction term of third
	order in the external potential, $A_\mu A_\nu\partial_\mu A_\nu$,
	vanishes for a static Coulomb potential.
\end{references}
\end{document}